\documentclass[letterpaper]{article}

\usepackage[T1]{fontenc}
 \usepackage{url}
\usepackage{geometry}
\usepackage{setspace}

\usepackage[numbers,sort&compress]{natbib}

\usepackage{graphicx}
\usepackage{float}
\newfloat{scheme}{htbp}{los}
\floatname{scheme}{Scheme}
\floatname{chart}{Chart}
\newfloat{graph}{htbp}{loh}

\usepackage{listings}
\usepackage[most]{tcolorbox}
\tcbuselibrary{listings,breakable}

\lstdefinestyle{promptstyle}{
  basicstyle=\ttfamily\small,
  columns=fullflexible,
  breaklines=true,
  breakatwhitespace=true,
  keepspaces=true,
  showstringspaces=false,
  frame=none
}

\usepackage{booktabs}
\usepackage{threeparttable}
\usepackage{siunitx}
\usepackage{makecell}

\usepackage{subcaption}
\usepackage{amsfonts}

\usepackage{booktabs}      % \toprule \midrule \bottomrule
\usepackage{threeparttable}% table notes
\usepackage{siunitx}       % decimal alignment

\usepackage[pdfencoding=auto]{hyperref}

\newtcolorbox{sciencebox}[2][]{%
  enhanced,
  breakable,
  colback=gray!10,
  colframe=black!60,
  boxrule=0.6pt,
  arc=0.8mm,
  left=6pt,right=6pt,top=6pt,bottom=6pt,
  title=\textbf{#2},
  fonttitle=\normalsize,
  coltitle=black,
  attach title to upper,
  #1
}
\usepackage{chemformula} % Formulas using \ch{}
\usepackage[version = 4]{mhchem} % Formulas using \ce{}

\usepackage{authblk}

\author[1]{Qiulin Zeng}
\affil[1]{Department of Math,
  University of Southern California,
  Los Angeles, CA, 90007}

\author[2]{Tahiya Chowdhury}
\affil[2]{Department of Computer Science,
  Colby College,
  Waterville, ME, 04901}

\author[3]{Md Shafayat Hossain\thanks{*Correspondence and requests for materials should be addressed to M.S.H. (shossain@seas.ucla.edu).}}

\affil[3]{Department of Materials Science and Engineering, University of California, Los Angeles, Los Angeles, CA 90095, United States}

\affil[3]{California NanoSystems Institute, University of California, Los Angeles, Los Angeles, CA 90095, United States}

\title{Generative Discovery of Magnetic Insulators under Competing Physical Constraints}
\date{*Correspondence to: shossain@seas.ucla.edu}

\begin{document}

\maketitle

\begin{abstract}
Discovering materials that must simultaneously satisfy multiple competing constraints remains a central challenge in computational materials design, particularly in data-scarce regimes where conventional data-driven approaches are least effective \cite{doi:10.1126/science.aat2663}. Magnetic insulators represent a stringent example: the electronic conditions that favor magnetic order often also promote metallicity, while insulating behavior suppresses the interactions that stabilize magnetism. As a result, experimentally viable magnetic insulators are rare and difficult to identify through conventional screening. Here, we introduce \textit{MagMatLLM}, a constraint-guided generative discovery framework that integrates language-model-based crystal generation with evolutionary selection, surrogate screening, and first-principles validation to target simultaneous stability, magnetism, and insulating behavior. Unlike stability-first approaches, the framework enforces functional constraints during generation and selection, steering the search toward sparsely populated regions of materials space defined by competing physical requirements. Using this workflow, we identify twelve previously unreported candidate magnetic insulators, including Tm$_4$Co$_2$Cr$_2$O$_{12}$ and Cr$_4$Nb$_2$O$_{12}$. Of these, ten are dynamically stable by phonon analysis and exhibit finite band gaps and nonzero magnetic moments in spin-polarized density functional theory calculations. Beyond the specific compounds identified here, this work establishes a general constraint-guided paradigm for multi-objective materials discovery in sparse chemical spaces and provides a transferable strategy for the design of quantum materials under competing physical constraints.
\end{abstract}

\section*{Keywords}

Magnetic Insulators, 
LLM-guided Materials Design, 
Crystal Structure Design, 
Genetic Algorithms, 
Density Functional Theory, 
Multi-objective Materials Optimization

\section{Introduction}

Magnetic insulators occupy a unique and increasingly important niche in modern materials science~\cite{Chumak2015Magnonics}. By combining long-range magnetic order with suppressed charge transport, they enable low-dissipation spin propagation, topological phases stabilized by magnetism, and emerging routes toward quantum and spintronic devices. Prototypical systems such as yttrium iron garnet~\cite{Uchida_2010} have long served as benchmark platforms for magnonics and spin transport, while magnetic topological insulators underpin quantum anomalous Hall physics and related topological phenomena~\cite{2019NatRP...1..126T,Wang2021Magnetic}. Despite their importance, experimentally viable magnetic insulators remain scarce beyond a few well-studied material families~\cite{Emori2021Ferrimagnetic,Huang2017Layer}.

The difficulty arises from an intrinsic incompatibility in electronic structure: magnetism typically emerges from partially filled electronic bands, whereas insulating behavior requires the absence of low-energy carriers. These competing requirements sharply restrict the accessible chemical space, leading to sparse datasets and strong chemical bias in known materials. In such regimes, conventional discovery approaches face inherent limitations. Density functional theory (DFT)-based high-throughput screening is computationally expensive and scales poorly with compositional complexity~\cite{Choudhary2021High}, while data-driven machine learning methods struggle due to limited training data and the absence of well-defined interpolation regimes~\cite{article}.

Recent advances in generative modeling and large language models (LLMs) offer a promising alternative pathway for materials discovery~\cite{gan2025largelanguagemodelsinnate,tshitoyan_dagdelen_weston_dunn_rong_kononova_persson_ceder_jain_2019}. LLM-based crystal generation frameworks can propose chemically plausible structures with minimal domain-specific training, especially when coupled with evolutionary search and downstream validation~\cite{gan2025largelanguagemodelsinnate}. However, existing approaches typically prioritize thermodynamic stability and apply functional constraints—such as magnetism or band gap—only as post hoc filters. This stability-first paradigm is inefficient for discovering rare materials defined by multiple competing criteria.

The discovery of magnetic insulators remains challenging despite advances in computational and data-driven methods. Conventional approaches—density functional theory (DFT) combined with chemical intuition—are computationally demanding and often biased toward well-explored chemistries. High-throughput platforms such as the Materials Project~\cite{Jain2013MaterialsProject,article,Ward_2016} and the Topological Magnetic Materials Database~\cite{Bradlyn_2017,Vergniory2019,doi:10.1126/science.abg9094} have significantly expanded accessible materials space, while machine learning (ML) has accelerated property prediction~\cite{deng_2023_chgnet,Xie_2018,deng2023chgnetpretraineduniversalneural}. Generative models, including crystal graph networks and diffusion-based frameworks~\cite{KUSABA2022111496,NOH20191370}, can now propose chemically plausible structures. However, progress remains limited in sparse-data regimes: high-quality datasets for magnetic topological materials are scarce, and most generative approaches lack mechanisms to efficiently optimize candidates under multiple competing objectives. Recent work suggests that large language models (LLMs) provide a promising alternative, acting as flexible crystal generators when combined with validation pipelines~\cite{gan2025largelanguagemodelsinnate,tshitoyan_dagdelen_weston_dunn_rong_kononova_persson_ceder_jain_2019}. However, existing implementations largely rely on one-shot generation or stability-first filtering, which limits their effectiveness for multi-objective materials discovery in chemically sparse regimes.

Here we address these limitations by introducing a \emph{constraint-first generative discovery framework}. By embedding physical constraints during generation rather than applying them only as post hoc filters, the framework increases the likelihood of identifying candidates that satisfy multiple competing objectives in sparse regimes. In this approach, functional objectives—magnetism, insulating behavior, and stability—are embedded directly into the generative and selection loop. Rather than generating large numbers of candidates and subsequently filtering them, candidate structures are iteratively proposed, evaluated, and refined under explicit physical constraints, enabling efficient exploration of sparse and highly constrained regions of materials space.

To implement this strategy, we develop \emph{MagMatLLM}, a language-model–guided evolutionary framework tailored for magnetic insulators. MagMatLLM couples crystal generation with multi-objective selection for stability, magnetism, and band gap within a closed-loop workflow, followed by first-principles validation using lattice dynamics and electronic-structure calculations. Using this framework, we identify twelve previously unreported candidate magnetic insulators, including Tm$_4$Co$_2$Cr$_2$O$_{12}$ and Cr$_4$Nb$_2$O$_{12}$. Of these, ten are dynamically stable by phonon analysis and exhibit finite band gaps and nonzero magnetic moments in spin-polarized density functional theory calculations. Beyond individual compounds, this work establishes a scalable computational strategy for multi-constraint materials discovery in data-scarce regimes and provides a transferable paradigm for constraint-guided design of quantum materials.

\section{Related Work}
Crystal structure generation (CSG) methods span several paradigms that balance physical fidelity, controllability, and computational cost.

Traditional approaches, including evolutionary algorithms (USPEX)~\cite{GLASS2006713} and particle swarm optimization (CALYPSO)~\cite{WANG20122063}, explore the global energy landscape without prior structural assumptions. While highly reliable, their reliance on repeated DFT relaxations makes them computationally expensive and difficult to scale across large chemical spaces.

Deep generative models—such as VAEs, GANs, and diffusion-based frameworks—offer a data-driven alternative, enabling rapid generation of candidate structures. For example, Cond-CDVAE achieves strong structural reconstruction accuracy under property conditioning~\cite{Ye_2024}. However, these models typically provide limited control over crystallographic symmetry and space-group distributions, often requiring additional filtering to ensure physical validity.

To address this, symmetry-aware methods incorporate crystallographic priors directly into the generation process. Approaches such as WyCryst~\cite{zhu2024wycrystwyckoffinorganiccrystal} and symmetry-conditioned generative models (e.g., SymmCDVAE~\cite{ishii2026symmetryawareconditionalgenerationcrystal}) improve structural fidelity by enforcing or guiding symmetry constraints. Nevertheless, their dependence on predefined symmetry descriptions can restrict flexibility, particularly for low-symmetry or disordered systems.

More recently, LLM-driven frameworks have emerged as a flexible alternative, leveraging pre-trained language models to generate crystal structures from natural language prompts. Methods such as \emph{MatLLMSearch}~\cite{gan2025largelanguagemodelsinnate} demonstrate that LLMs can achieve competitive efficiency without domain-specific training. However, current implementations remain largely unconstrained, with outputs sensitive to prompt design and lacking explicit integration of crystallographic or multi-objective criteria~\cite{jia2024llmatdesignautonomousmaterialsdiscovery}.

\paragraph{Positioning of this work.}
Across these approaches, a common limitation is that functional properties are typically imposed \emph{after} generation, rather than embedded within it. This is particularly restrictive for quantum materials, where properties such as magnetism and insulating behavior arise from competing physical constraints. Here, we address this gap by introducing a constraint-guided, closed-loop CSG framework that integrates generation, evaluation, and selection. By combining LLM-based generation with evolutionary optimization and multi-objective screening, our approach enables direct exploration of materials spaces defined by simultaneous stability and functionality, even in data-scarce regimes.

\section{Methodology behind MagMatLLM Architecture}

\begin{figure}
    \centering
    \includegraphics[width=0.95\linewidth]{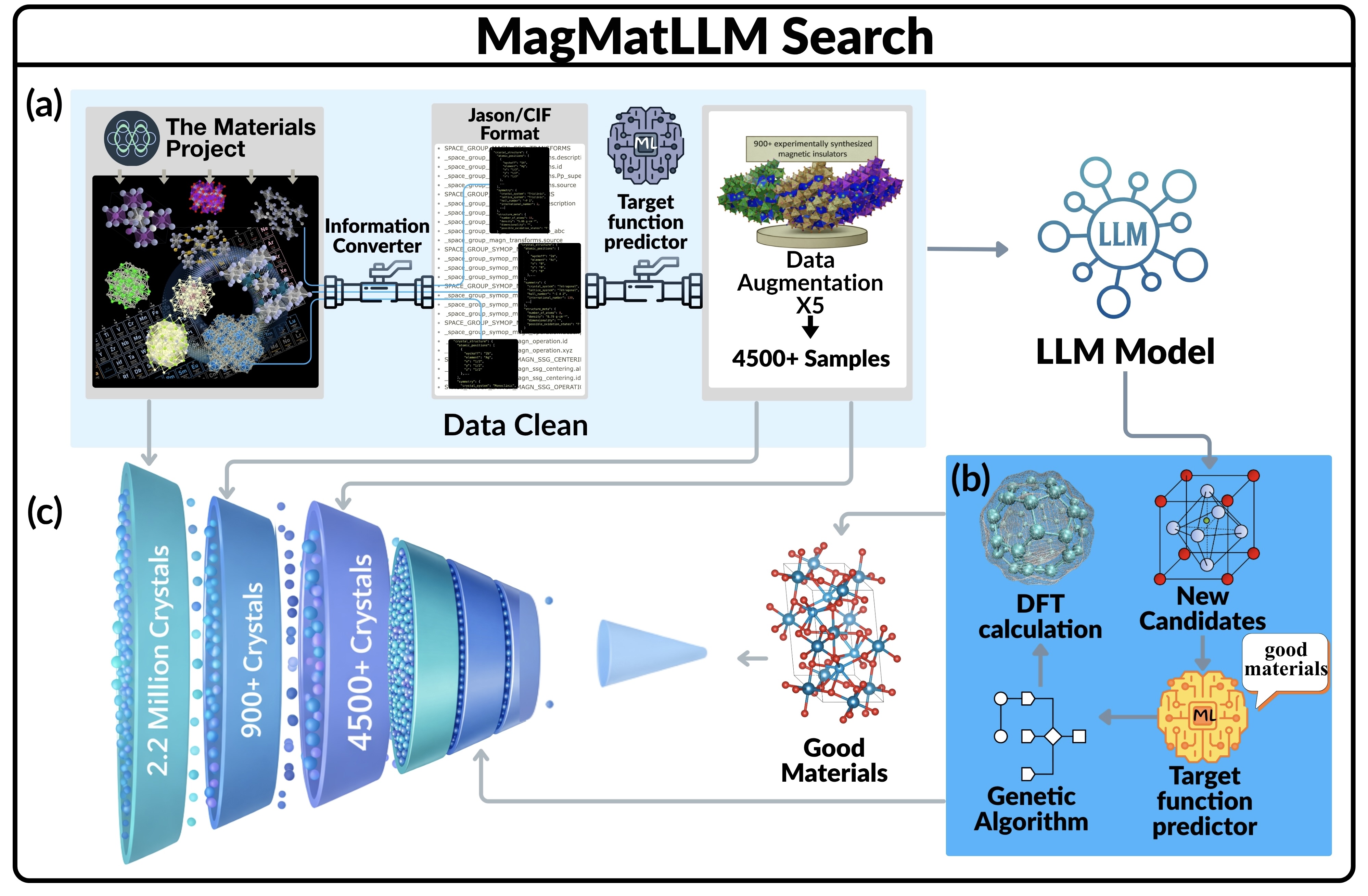}
   \caption{\textbf{MagMatLLMSearch Framework.}
Schematic illustration of the large language model (LLM)-assisted crystal materials discovery framework. (a) Crystal structure data from The Materials Project are converted to standardized formats (e.g., JSON/CIF), cleaned, and augmented to generate over 4,500 samples for model training. (b) The trained LLM target function predictor is combined with a genetic algorithm and DFT calculations to generate and evaluate new candidate materials. (c) (c) A funnel-based screening strategy progressively narrows the search space from 2.2 million crystals to 900+ crystals and 4,500+ crystals, ultimately identifying high-quality candidate materials. The funnel highlights how constraint-guided screening concentrates the search toward functionally relevant candidates.}

    \label{fig:MagMatLLMSearch}
\end{figure}

MagMatLLM is a constraint-guided, closed-loop materials discovery framework that integrates large language model (LLM)–based crystal generation with evolutionary search, multi-objective screening, and first-principles (see Fig.~\ref{fig:MagMatLLMSearch}). Unlike prior LLM-driven crystal generation approaches that primarily optimize for thermodynamic stability~\cite{gan2025largelanguagemodelsinnate}, our framework embeds multiple functional objectives—magnetism, insulating behavior, and stability—directly into the generation and selection loop. This \emph{constraint-first} design ensures that candidate materials are iteratively refined toward target properties rather than filtered only after large-scale generation.

The workflow consists of three tightly coupled stages: (i) LLM-guided crystal structure generation, (ii) multi-stage screening using physical constraints and surrogate models, and (iii) iterative evolutionary selection. Together, these components form a closed-loop optimization process that progressively drives the search toward materials satisfying multiple competing criteria.

\subsection{LLM-guided Materials Search}

LLM guided materials search approaches, e.g., \emph{MatLLMSearch}~\cite{gan2025largelanguagemodelsinnate}, demonstrate that general-purpose LLMs, when combined with evolutionary search, can generate large numbers of thermodynamically stable or metastable crystal structures without any domain-specific fine-tuning~\cite{gan2025largelanguagemodelsinnate}. This framework treats materials as a dynamic population, evolving through repeated \emph{generation → evaluation → selection} cycles. Only the most promising material candidates advance to the next stage, the expensive but definitive DFT verification, enabling an efficient closed-loop design workflow.

\subsubsection{Initialization}
An initial stable pool $D$ of crystal structures serves as the parent population, optionally supplemented by a reference pool 
$R$ to increase diversity. Parent groups are sampled from $D$ and provided to the LLM as exemplars for guided generation.

\subsubsection{Reproduction}
The LLM acts as a generative engine that proposes new candidate structures by modifying lattice parameters, atomic coordinates, elemental compositions, or structural motifs of the parent materials. These transformations are guided by prompts that enforce chemical plausibility and crystallographic validity while encouraging exploration of nearby compositional and structural space~\cite{gan2025largelanguagemodelsinnate}.

\subsubsection{Evaluation}

The candidate structures are evaluated through a two-stage screening procedure. 
The first stage enforces basic physical validity, while the second stage assesses 
key material properties using machine-learning-based predictors.

\paragraph{Stage I: Basic Validity Checks}
In the first stage, candidate structures are filtered according to a set of basic physical validity rules. 
All structures must satisfy three-dimensional periodic boundary conditions to ensure a valid crystalline lattice. Unrealistic atomic geometries are removed by enforcing minimum interatomic distance constraints. 
For any atomic pair $i,j$, the interatomic distance $d_{ij}$ must satisfy

\begin{equation}
d_{ij} \ge \alpha (r_i + r_j)
\label{eq:distance_constraint}
\end{equation}

where $r_i$ and $r_j$ denote the covalent radii of the corresponding elements and $\alpha$ is a tolerance factor 
(typically $0.7$--$0.8$). Structures violating this condition are rejected due to atomic overlap or unrealistically 
short bond lengths. Charge neutrality is also enforced such that the total oxidation-state-weighted charge within each unit cell equals zero. Structures that violate charge neutrality are discarded.

In addition, structures containing isolated atoms without neighbors within a reasonable coordination radius are removed. 
Finally, duplicate structures are identified using structural fingerprints derived from lattice parameters and atomic 
coordinates, and removed to prevent redundant downstream evaluation.

\paragraph{Stage II: ML-based Property Evaluation}
Structures that pass the initial screening are further evaluated using machine-learning models to assess their thermodynamic and mechanical properties. 
First, rapid geometry optimization is performed using \texttt{CHGNet} \cite{deng_2023_chgnet}to obtain relaxed atomic configurations. 
The formation energy ($E_d$), which represents the energy required to dissociate a material into its constituent elements, is then calculated. 
Negative values of $E_d$ indicate thermodynamic stability, whereas values in the range of $0$--$0.1$~eV/atom correspond to metastable candidates. 
Next, the distance to the thermodynamic convex hull ($E_c$) is evaluated, with values below $0.05$~eV / atom typically considered stable. 
In addition, the bulk modulus is estimated using \texttt{CHGNet}, with larger values indicating greater mechanical robustness. These surrogate evaluations provide a computationally efficient approximation of material properties, enabling large-scale screening prior to first-principles validation.

\subsubsection{Selection}

At each generation, parent structures $D$, newly generated candidates, and optional reference structures $R$ are merged into a unified candidate pool. Candidates are ranked according to the optimization objective, which prioritizes thermodynamic stability while incorporating mechanical robustness and magnetic properties. The top $K \times P$ candidates, where $K$ is the population size and $P$ is the number of parents per group, are selected to form the parent pool for the next generation. This iterative selection mechanism drives progressive improvement in candidate quality across generations.

\subsection{Algorithm Design}

\begin{figure}
    \centering
    \includegraphics[width=0.70\linewidth]{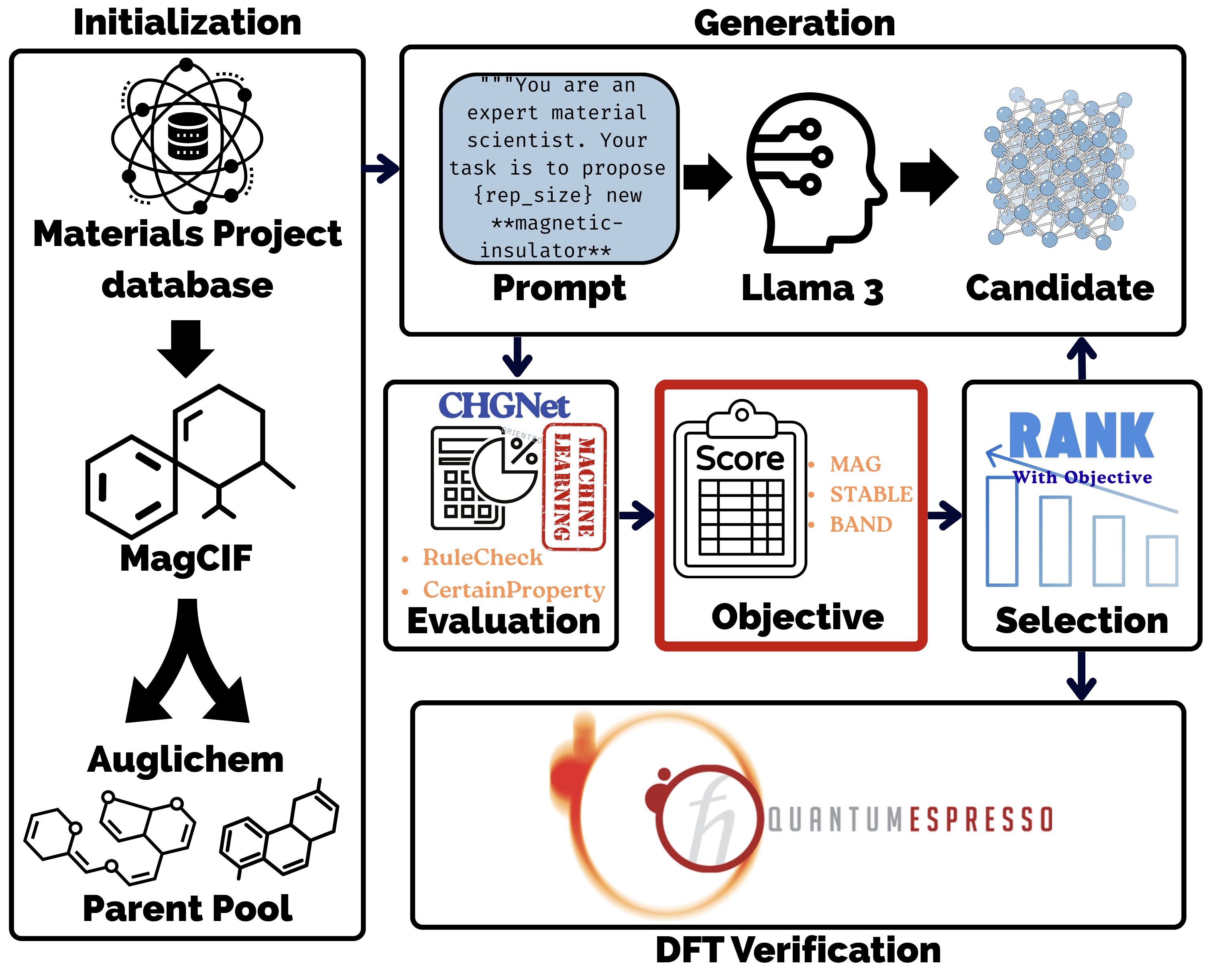}
   \caption{\textbf{LLM-driven materials search loop with objective scoring.}
Light-gray boxes depict the original LLM materials search loop (Initialization → Reproduction → Selection → Final DFT).
Our additions are the red modules: \emph{Objective} (normalizing $E_d$, $M_{\mathrm{tot}}$, and $b$ to rank candidates),
which unified multi-criteria selection.}

    \label{fig:placeholder}
\end{figure}

\emph{MatLLMSearch} framework~\cite{gan2025largelanguagemodelsinnate} is optimized primarily for thermodynamic stability. In contrast, practical materials discovery often requires simultaneous optimization of multiple functional properties. In this work, we focus on magnetic insulators, which are defined by a combination of magnetic ordering, finite electronic band gap, and thermodynamic stability. The overall LLM-driven search loop and the integration of multi-objective scoring are illustrated in (see Fig.~\ref{fig:placeholder}).

\subsubsection{Design Rationale}

The magnetic property of each candidate is quantified by its total magnetic moment $M_{\mathrm{tot}}$, where $M_{\mathrm{tot}} > 0$ indicates net magnetic ordering. The insulating behavior is characterized by the electronic band gap $E_g$, and candidates with $E_g \ge 0.025~\mathrm{eV}$ are retained to exclude metallic systems~\cite{PBE1996}. Material stability is assessed using complementary indicators, including the bulk modulus $b$ and the distance to the convex hull $E_d$.

A material satisfying $M_{\mathrm{tot}} > 0$ and $E_g \ge 0.025~\mathrm{eV}$ is classified as a magnetic insulator in this work. This operational definition is used consistently at the PBE level for screening and benchmarking. These criteria are enforced throughout the iterative generation and selection process, ensuring that the search remains focused on the target material class.

Candidate structures are first proposed by the LLM and subsequently screened using surrogate predictors. 
The highest-scoring structures are then retained as parents for the next generation, enabling progressive optimization toward the desired magnetic insulating characteristics.

%As a first step, it is necessary to define the fundamental characteristics of such materials. The magnetic property is quantified by the total magnetic moment $M_{\mathrm{tot}}$, where $M_{\mathrm{tot}} > 0$ indicates a net magnetic ordering. The insulating property is characterized by the electronic band gap $E_g$; in this study, we adopt the criterion $E_g \geq 3 \ \mathrm{eV}$, corresponding to extremely low electrical conductivity. A material satisfying $M_{\mathrm{tot}} > 0$ and $E_g \geq 3 \ \mathrm{eV}$ is classified as a magnetic insulator for the purposes of this work. In addition to these functional properties, we also retain stability-related quantities: the bulk modulus $b$ and the distance to the convex hull $E_d$.

%The first step in our approach is to upload an mcif file containing magnetic data to the LLM, accompanied by a prompt to generate candidate structures. For a candidate structure $x$, we denote by $E_d(x)$ the distance to the convex hull (smaller values indicate greater thermodynamic stability), by $b(x)$ the bulk modulus (larger values are preferable), and by $M_{\mathrm{tot}}(x)$ the total magnetic moment (larger values indicate stronger magnetic ordering). In generation $i$, a set of parents $P_i$ produces children $C_i$ through LLM-guided reproduction. All candidates are then scored, and the next set of parents $P_{i+1}$ is selected according to the defined criteria.

\subsubsection{Pre-processing and Normalization}
To place heterogeneous objectives on a comparable scale, we first apply winsorization \cite{Dixon1974} to the formation energy $E_d$, capping extreme values at the 95th percentile to prevent a few highly unstable candidates from dominating the scale. Formally,
\begin{equation*}
  \widetilde{E}_d(x)=\min\{E_d(x),\,E_{d,95}\},
\end{equation*}
This ensures that large $E_d$ values (corresponding to unrealistic structures) do not distort normalization. We then apply min–max normalization within each generation. \footnote{\(\varepsilon>0\) prevents division by zero here}.

\begin{equation*}
  E_d^{\star}(x)=\frac{\widetilde{E}_d(x)-E_{d,\min}}{\max(E_{d,\max}-E_{d,\min},\,\varepsilon)},\quad
  b^{\star}(x)=\frac{b(x)-b_{\min}}{\max(b_{\max}-b_{\min},\,\varepsilon)},\quad
\end{equation*}
\begin{equation*}
  M_{\mathrm{tot}}^{\star}(x)=\frac{M_{\mathrm{tot}}(x)-M_{\min}}{\max(M_{\max}-M_{\min},\,\varepsilon)}
\end{equation*}

Note that for a candidate structure $x$, smaller $E_d(x)$ values \textbf{(distance to the convex hull)} indicate greater thermodynamic stability), and larger $b(x)$ values \textbf{(bulk modulus) }and larger $M_{\mathrm{tot}}(x)$ values \textbf{(the total magnetic moment)} indicate mechanical robustness and stronger magnetic ordering, which are preferred. 
Hence, when aggregating the properties, we invert polarity so that all objectives follow the rule \textit{smaller objective means better performance}:
\begin{equation*}
  \overline{b}(x)=1-b^{\star}(x),\qquad
  \overline{M}_{\mathrm{tot}}(x)=1-M_{\mathrm{tot}}^{\star}(x).
\end{equation*}

\begin{figure}[H]
  \centering
    \includegraphics[width=\linewidth]{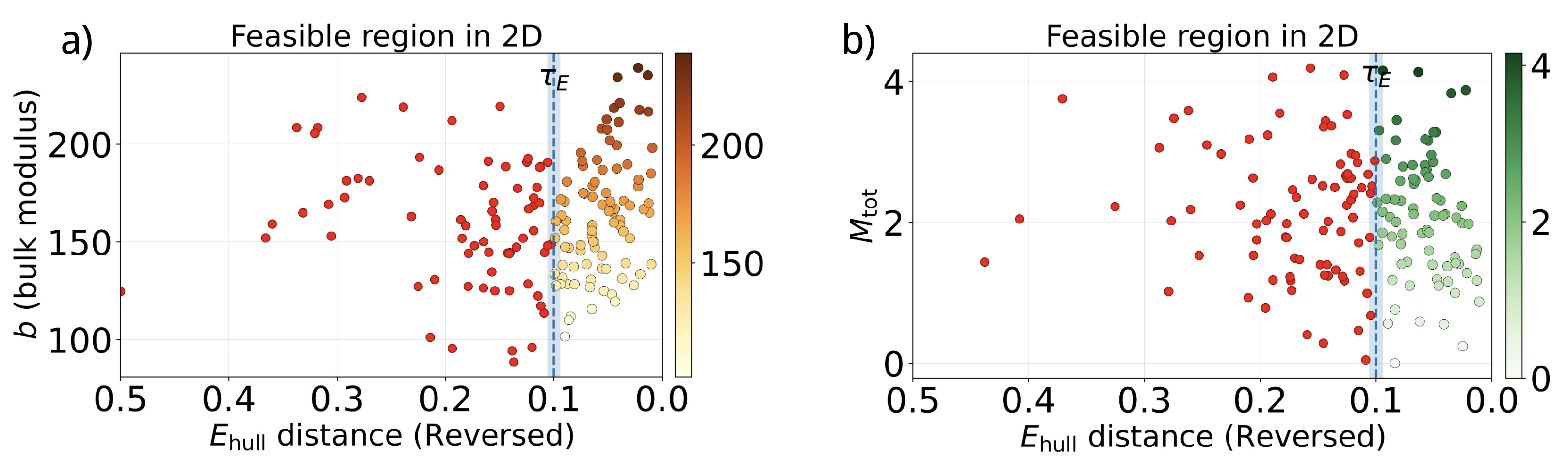}
  \caption{\textbf{Feasible regions in two-dimensional objective spaces.} 
(a) Distribution of candidates in reversed convex hull distance and bulk modulus. 
(b) Distribution in reversed convex hull distance and total magnetic moment. 
The horizontal axis shows the reversed convex hull distance, where smaller $E_{\mathrm{hull}}$ values 
(i.e., points toward the right side) correspond to more thermodynamically stable structures. 
The dashed vertical line indicates the stability threshold $\tau_E$. 
Candidates located to the right of this threshold ($E_{\mathrm{hull}} \le \tau_E$) constitute the feasible region, 
while those on the left are considered unstable. 
Red points denote candidates that fail to satisfy the stability constraint, whereas the colored points in the feasible 
region represent candidates that meet the stability criterion. 
The color intensity encodes the normalized objective value, with darker colors indicating better overall performance 
in terms of stability, mechanical robustness, and magnetic properties.}
\label{fig:feasible_region}
\end{figure}

For any two candidates $x_1,x_2$, if, with all other components equal, $b(x_1) > b(x_2)$ (or $M_{\mathrm{tot}}(x_1) > M_{\mathrm{tot}}(x_2)$), then
\[
  O_{\mathrm{ws}}(x_1)\le O_{\mathrm{ws}}(x_2).
\]
Therefore, inverting the polarity does not alter the underlying physical nature; it merely improves consistency in the numerical implementation and the optimization procedure. The resulting normalized objective space and the feasible regions defined by stability constraints are illustrated in Fig.~\ref{fig:feasible_region}.

\subsubsection{Optimization Objective}

Our objective function is utilized to search for crystals with specific properties and stability (in this case, we are looking for magnetic insulators). We consider two complementary strategies for this goal.

\textbf{(I) Weighted Sum (Continuous Trade-off).} Considering nonnegative weights \(\alpha_E,\alpha_b,\alpha_m\) for $E_d$, $b$, and $M_{tot}$, we can describe the objective function as follows:
\begin{equation}
  \label{eq:ws}
  O_{\mathrm{ws}}(x)=\alpha_E\,E_d^{\star}(x)+\alpha_b\,\overline{b}(x)+\alpha_m\,\overline{M}_{\mathrm{tot}}(x),
\end{equation}
Here, smaller \(O_{\mathrm{ws}}\) values indicate better overall trade-offs among stability, mechanical robustness, and magnetic properties. This method is useful when all objectives are reasonably reliable and no strict thresholds are required.

\textbf{(II) Lexicographic (Threshold-first).}
Given a stability threshold \(\tau_E\), a penalty \(\lambda>0\), and coefficients \(\beta_E\!\ll\!1\), \(\beta_b,\beta_m>0\),
\begin{equation}
  \label{eq:lex}
  O_{\mathrm{lex}}(x)=
  \begin{cases}
    \beta_E\,E_d(x)-\beta_b\,b(x)-\beta_m\,M_{\mathrm{tot}}(x), & \text{if } E_d(x)\le \tau_E,\\[4pt]
    E_d(x)+\lambda, & \text{if } E_d(x)>\tau_E~.
  \end{cases}
\end{equation}

%This implements “\emph{stability first, then maximize \(b\) and \(M_{\mathrm{tot}}\)}”.

Here, stability is enforced as a hard constraint ($E_d \leq \tau_E$), which prioritizes ensuring feasibility in early searches or when surrogate models are noisy (\textit{stability first, then optimize $b$ and $M_{tot}$}).

\textbf{Intuition.} When objectives can be balanced against each other, there are no non-negotiable `red lines.' Suppose each indicator (e.g., bulk modulus, magnetic moment, etc.) is provided by a relatively reliable proxy model. In that case, Objective I with weighted sum can enable continuous trade-offs between stability and performance, making it suitable for fine-tuning and exploring the Pareto frontier in the later stages. However, if there are hard thresholds that must be met first (e.g., $Eg \geq 0.025 eV$, $M_{tot} > 0$) or when the surrogate models have significant noise and feasible solutions are scarce in the early stages, Objective II, lexicographic with threshold-based, should be used. In summary, Objective I supports smooth exploration of the Pareto frontier, while II enforces `stability first, then optimizes properties.' We employ them adaptively depending on the data quality and the search stage.

\subsection{Candidate Ranking and Parent Update}
At the end of each generation, we retain only stable or metastable magnetic candidates with small band gap, defined by the screening thresholds $E_d(x) \leq 0.1~\mathrm{eV/atom}$, $M_{\mathrm{tot}}(x) > 0$, and $E_g \ge 0.025~\mathrm{eV}$. This ensures that unstable structures are filtered out early, avoiding redundant computation. To maintain diversity, the pool is optionally augmented with a reference set $R$:
\begin{equation}
  \label{eq:pool}
  \mathcal{U}_i=\bigl\{\,x\in P_i\cup C_i\cup R \;\big|\; E_d(x)\le 0.1~\mathrm{eV/atom}\,\bigr\}.
\end{equation}

Let \(L=K \times P\) be the target parent count. 
The candidates in $\mathcal{U}_i$ are ranked by the chosen objective function---either the weighted-sum or $O_{\mathrm{ws}}$ (Eq.~\ref{eq:ws}) or lexicographic objective $O_{\mathrm{lex}}$ (Eq.~\ref{eq:lex}). 
The top $L$ candidates are retained, and new parent groups are sampled from this set for the next generation:
%Using either \(O_{\mathrm{ws}}\) (Eq.~\eqref{eq:ws}) or \(O_{\mathrm{lex}}\) (Eq.~\eqref{eq:lex}) as the scoring function \(O(\cdot)\), we sort \(\mathcal{U}_i\) in ascending order of \(O\) and keep the best \(L\) items:
\begin{equation}
  \label{eq:select}P_{i+1}=\operatorname{Sample}\!\Big(\operatorname{Top}_{L}\big(\mathcal{U}_i;\,O(\cdot)\big)\Big),
\end{equation}
where \(\operatorname{Top}_{L}(\cdot;\,O)\) returns the \(L\) candidates with the smallest objective value, and \(\operatorname{Sample}(\cdot)\) forms the modified parent groups for the next generation. 

\section{Experimental Results}

\begin{table}[htbp]
\centering
\small
\caption{Representative MagMatLLM candidates passing surrogate screening. Objective: multi-criteria score (lower is better). Formation energy (eV/atom): stability indicator (more negative is more stable). $E_{\mathrm{hull}}$ (eV/atom): energy above the convex hull (smaller is better). Magmom ($\mu_B$): total magnetic moment per formula unit.}

\label{tab:good-materials}
\begin{tabular}{l S[table-format=1.3] S[table-format=1.2] S[table-format=1.3] S[table-format=2.3]}
\toprule
Composition & {Objective} & {Formation energy (eV/atom)} & {E$_{\mathrm{hull}}$ (eV/atom)} & {Magmom ($\mu_B$)} \\
\midrule
Co$_{2}$Si$_2$Tb$_1$ & 0.124 & -4.11 & 0.032 & 3.170 \\
Yb$_3$ & 0.150 & -3.60 & 0.053 & 0.110 \\
Cr$_4$Nb$_2$O$_{12}$ & 0.073 & -8.74 & 0.039 & 15.620 \\
Mn$_2$O$_8$W$_2$ & 0.091 & -8.43 & 0.031 & 6.955 \\
Nd$_1$Mn$_2$Si$_2$ & 0.021 & -2.16 & 0.031 & 8.230 \\
Y$_3$ & 0.210 & -3.82 & 0.059 & 0.320 \\
Sm$_4$Mn$_2$Ir$_2$O$_{12}$ & 0.117 & -8.32 & 0.055 & 7.310 \\
Co$_2$Si$_2$Y & 0.280 & -7.04 & 0.024 & 0.130 \\
Hf$_2$Si$_4$ & 0.300 & -5.38 & 0.038 & 0.160 \\
Ho$_1$B$_2$O$_6$V$_1$ & 0.260 & -8.74 & 0.052 & 1.920 \\
Tm$_4$Co$_2$Cr$_2$O$_{12}$ & 0.051 & -8.47 & 0.018 & 8.460 \\
Gd$_1$Mn$_2$Si$_2$ & 0.170 & -7.04 & 0.024 & 6.120 \\
\bottomrule
\end{tabular}
\end{table}

\subsection{Stability Rate}
We first evaluate the ability of the proposed framework to generate thermodynamically viable candidate materials.

For the baseline framework \emph{MatLLMSearch}, $1{,}479$ candidate materials are generated, of which \textbf{78.38\%} are classified as metastable ($E_{\mathrm{hull}} < 0.1~\mathrm{eV/atom}$) using \texttt{CHGNet} evaluation, representing a substantial improvement over earlier LLM-based generators such as \emph{CrystalTextLLM-70B} ($49.8\%$)~\cite{gruver2025finetunedlanguagemodelsgenerate}. Furthermore, \textbf{31.7\%} of candidates are confirmed to be thermodynamically stable ($E_{\mathrm{hull}} < 0$) by DFT validation.

For the proposed \emph{MagMatLLM} framework, a total of 82 unique and previously unreported candidate materials are generated from 1{,}000 parent structures. Among these, 12 candidates are predicted to be thermodynamically stable by machine-learning-based evaluation (Table~\ref{tab:good-materials}), corresponding to a surrogate-level stability rate of $\tfrac{12}{82} \approx 14.6\%$.

\subsection{Benchmarking: Comparison of Different Approaches}
\label{sec:benchmark}

% Requires: \usepackage{booktabs}
% Optional: \usepackage{threeparttable}

Because the practical goal is \emph{discovery efficiency}—how often a method produces candidates that jointly satisfy stability, magnetism, and insulation under fixed compute—we benchmark all methods using identical screening and DFT validation criteria. We report both stage-wise pass rates and the joint magnetic-insulator success rate, together with GPU-hours per 1,000 candidates to quantify cost-normalized yield.

\paragraph{Unified evaluation protocol.}
We benchmark all methods with identical surrogate screening and DFT validation rules.
Each generated candidate is processed by the same (i) structure relaxation / surrogate predictor, (ii) stability filter, and (iii) de-duplication pipeline.
We report stage-wise pass rates and the joint success rate:
\[
\text{MI (Magnetic Insulator) success} :=
\mathbf{1}[\text{Surrogate-stable}]
\wedge
\mathbf{1}[M_{\mathrm{tot}} > 0]
\wedge
\mathbf{1}[E_g \ge E_g^{\mathrm{th}}].
\]
To ensure reproducibility, we additionally report \textbf{GPU-hrs/1k} (accelerator hours per 1,000 candidates).

\paragraph{Variables control.}
We compare MagMatLLM against: (i) MatLLMSearch as the primary baseline, and (ii) a prior LLM generator baseline (CrystalTextLLM) when available.
To isolate the effect of \emph{prompting} from algorithmic changes, we include a \textbf{prompt-only MatLLMSearch} control:
we keep the original MatLLMSearch loop unchanged, and only replace its free-form prompt with our magnetic-insulator prompt schema (Box~1), i.e., explicit constraints + strict JSON output + per-site \texttt{magmom} fields.
MagMatLLM further introduces multi-objective ranking and selection beyond prompt engineering.

\begin{table}[H]
\centering
\begin{threeparttable}
\caption{Benchmark comparison on magnetic-insulator discovery under identical surrogate screening and DFT validation protocols.}
\label{tab:benchmark_mi}
\footnotesize
\setlength{\tabcolsep}{6pt}
\begin{tabular}{lccccccc}
\toprule
Method
& \makecell[c]{DFT-\\stable (\%)}
& \makecell[c]{Magnetic\\(\%)}
& \makecell[c]{Insulating\\(\%)}
& \makecell[c]{MI success\\(\%)}
& \makecell[c]{GPU-hrs\\ / 1k} \\
\midrule
CrystalTextLLM
& 10.8
& 42.2
& 8.1
& 2.1
& 450 \\

MatLLMSearch
& 25.7
& 55.9
& 40.3
& 8.3
& 360 \\

\textbf{MagMatLLM}
& 14.6
& 92.7
& 88.4
& 12.0
& 120 \\
\bottomrule
\end{tabular}

\begin{tablenotes}[flushleft]\footnotesize
\item \textbf{Definitions.}
DFT-stable: candidates have no imaginary frequency in Lattice-dynamics phonon dispersion which confirm dynamical stability.
Magnetic: fraction with nonzero total moment ($M_\mathrm{tot}>0$).
Insulating: fraction with band gap above threshold (use your paper’s threshold, e.g., $E_g\ge 0.025$~eV).
MI success: fraction satisfying \emph{all} required constraints jointly (DFT-stable $\wedge$ Magnetic $\wedge$ Insulating).
GPU-hrs/1k: total accelerator hours consumed per 1,000 generated candidates.
\end{tablenotes}
\end{threeparttable}
\end{table}

\begin{table}[H]
\centering
\begin{threeparttable}
\caption{Compute and scalability comparison (normalized throughput and cost).}
\label{tab:compute_scalability}
\footnotesize
\setlength{\tabcolsep}{7pt}
\begin{tabular}{lcccc}
\toprule
Method
& \makecell[c]{LLM\\params in billions}
& \makecell[c]{A100s\\used}
& \makecell[c]{Candidates\\per 24h}
& \makecell[c]{GPU-hrs\\ / 1k} \\
\midrule
MatLLMSearch\textit{(prompt-only)}
& 70B
& N
& $\sim$130
& $24\times N \times \frac{1000}{130}$ \\

\textbf{MagMatLLM} 
& 8B
& 1
& $\sim$200
& $24\times 1 \times \frac{1000}{200}=120$ \\
\bottomrule
\end{tabular}

\begin{tablenotes}[flushleft]\footnotesize
\item \textbf{Normalization.} We report throughput as generated candidates per 24 hours under the stated GPU setup.
GPU-hrs/1k is computed as
\[
\text{GPU-hrs/1k} = 24 \times (\#\text{GPUs}) \times \frac{1000}{\text{candidates/24h}}.
\]
Set $N$ to the actual number of A100 GPUs used for MatLLMSearch runs (e.g., 2 or 4), so the GPU-hrs/1k is fully reproducible.
\end{tablenotes}
\end{threeparttable}
\end{table}

\paragraph{Comparison.}
Tables \ref{tab:benchmark_mi} and \ref{tab:compute_scalability} show that \emph{MagMatLLM} achieves the highest joint magnetic-insulator success rate under identical screening and validation criteria while operating at lower computational cost. 
Although its overall DFT-stable rate is not the highest among the compared methods, \emph{MagMatLLM} exhibits dramatically higher pass rates for both magnetic and insulating constraints, leading to the best joint \emph{MI success} under a unified evaluation protocol. 
This indicates that \emph{MagMatLLM} does not simply improve structural viability, but instead biases generation toward the joint target region defined by stability, magnetism, and insulating behavior.

Importantly, these gains are achieved with significantly lower computational cost.
As shown in Table \ref{tab:compute_scalability}, \emph{MagMatLLM} operates with an 8B model on a single A100 GPU, yielding the lowest GPU-hours per 1,000 candidates.
When normalized by compute, \emph{MagMatLLM} attains the highest MI yield per GPU-hour, highlighting its superior end-to-end efficiency and scalability.
The prompt-only MatLLMSearch control further confirms that these improvements cannot be attributed to prompting alone, but arise from the proposed multi-objective ranking and selection strategy beyond prompt engineering.

\begin{figure}
    \centering
    \includegraphics[width=0.85\linewidth]{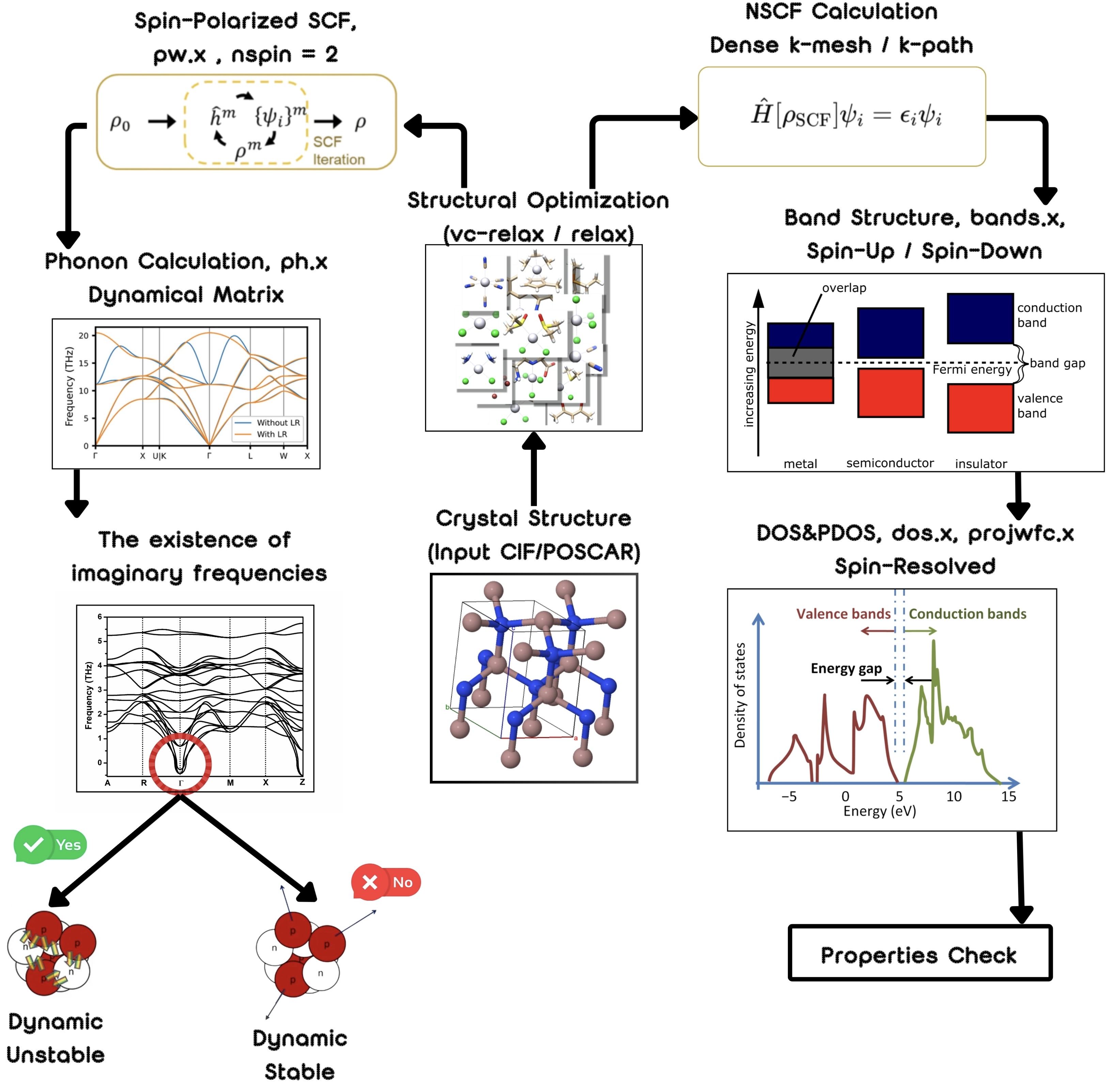}
   \caption{\textbf{Computational workflow for spin-polarized DFT calculations including dynamical stability and electronic structure analysis.}
Schematic workflow of the spin-polarized density functional theory (DFT) calculations performed using Quantum ESPRESSO. Starting from the crystal structure (input CIF/POSCAR), structural optimization (vc-relax/relax) is carried out, followed by spin-polarized self-consistent field (SCF) calculations (pw.x, nspin = 2). The converged charge density is used for two subsequent branches: (i) phonon calculations (ph.x) to evaluate the dynamical matrix and identify possible imaginary frequencies for stability assessment, and (ii) non-self-consistent field (NSCF) calculations with dense k-mesh or high-symmetry k-path to obtain spin-resolved band structures (bands.x) and density of states (DOS/PDOS, dos.x and projwfc.x). The workflow enables systematic evaluation of dynamical stability, band gap, exchange splitting, and electronic properties.}
\label{DFTworkflow}
\end{figure}

\paragraph{First-principles characterization.}
As part of the \emph{automated characterization and evaluation} stage of the design cycle, 
we performed three first-principles-based calculations to further assess the physical validity 
and electronic properties of each candidate material. 
First, lattice-dynamics phonon dispersion calculations are conducted to confirm the dynamical stability by verifying the absence of imaginary phonon frequencies across the Brillouin zone. 
Second, spin-polarized electronic band structures are computed to quantify fundamental band gaps and to assess insulating behavior through band dispersion. 
Finally, the total density of states (DOS) is analyzed to verify the magnitude of the electronic gap and to probe magnetic characteristics through spin splitting and orbital contributions.

\paragraph{\textbf{MagMatLLM} candidate selection.}
Due to the manageable scale of the candidate set, we performed first-principles calculations for all 12 materials that were predicted to be thermodynamically stable by our machine-learning surrogate model. Among these candidates, 10 were found to exhibit dynamical stability, as evidenced by the absence of imaginary phonon frequencies throughout the Brillouin zone. This high confirmation rate further supports the reliability of the surrogate-based screening stage. In the following sections, we present a detailed first-principles analysis of these ten dynamically stable materials, focusing on their structural, vibrational, and electronic properties. Additional representative candidates and their corresponding phonon dispersions and electronic structures are shown in Fig. S1 (see Supplemental Materials).

%For each material, we conducted three principal first-principles calculations: (i) lattice-dynamics (phonon) computations to assess dynamical stability via the absence of imaginary frequencies across the Brillouin zone; (ii) spin-polarized electronic-structure calculations to obtain the band dispersion and quantify the fundamental band gap; and (iii) total density-of-states (DOS) analyses to corroborate the gap magnitude and interrogate magnetic character through spin splitting and orbital contributions. The corresponding computational results are provided alongside and will be analyzed in detail in the following subsection.

\begin{figure}[htbp]
    \centering
    \includegraphics[width=0.95\textwidth]{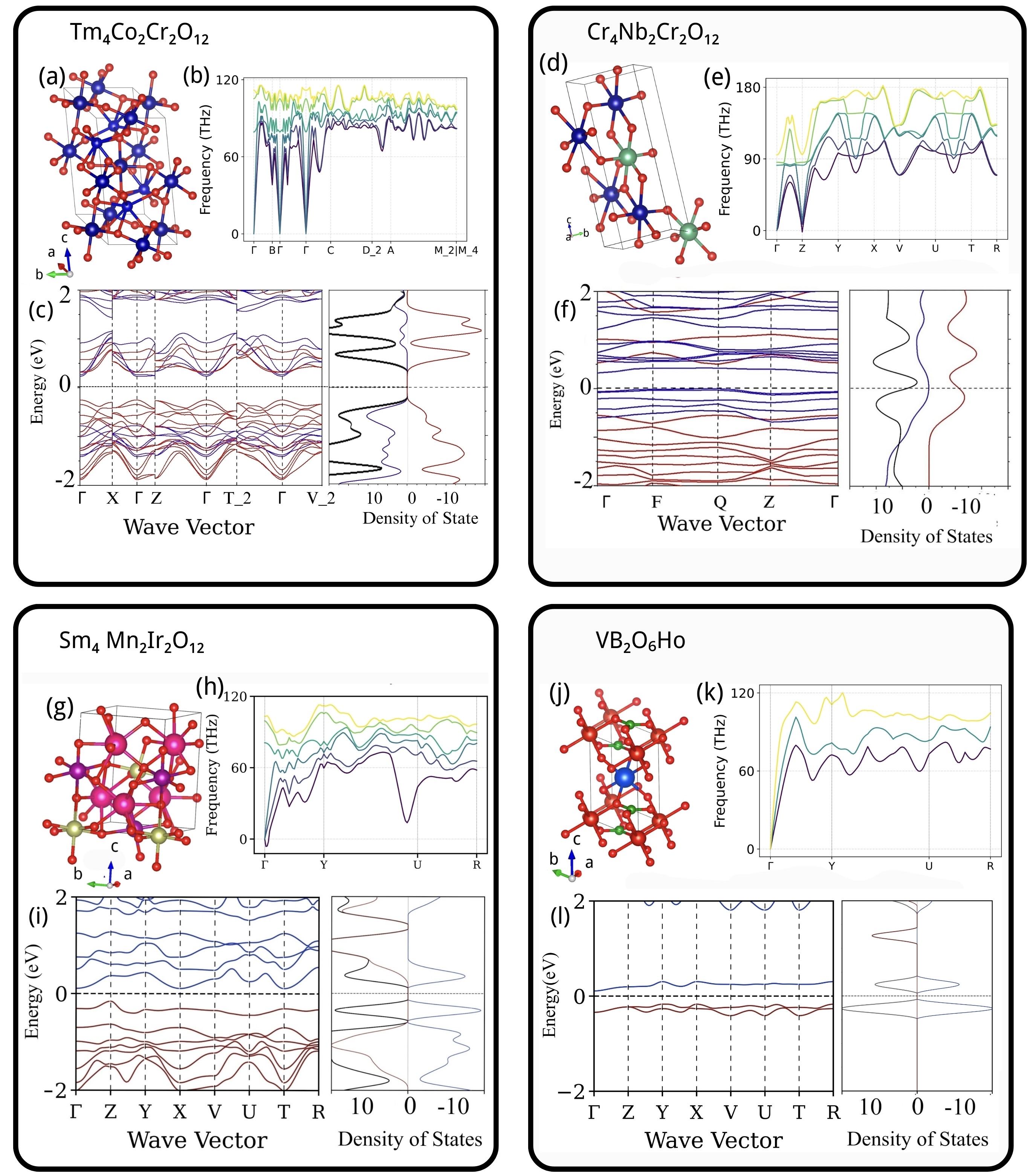}
    \caption{\textbf{Lattice dynamics and electronic structure of representative candidate materials.} Crystal structures of $\mathrm{Tm_4Co_2Cr_2O_{12}}$, $\mathrm{Cr_4Nb_2Cr_2O_{12}}$, $\mathrm{Sm_4Mn_2Ir_2O_{12}}$, and $\mathrm{B_2HoO_6V}$ are shown in panels \textbf{(a)}, \textbf{(d)}, \textbf{(g)}, and \textbf{(j)}, respectively (ball-and-stick representation with crystallographic axes). Phonon dispersions along high-symmetry $q$ paths are shown in \textbf{(b)}, \textbf{(e)}, \textbf{(h)}, and \textbf{(k)}, where the absence of imaginary modes confirms dynamical stability at 0\,K. Spin-polarized electronic band structures and the corresponding density of states (DOS) are presented in \textbf{(c)}, \textbf{(f)}, \textbf{(i)}, and \textbf{(l)}, with the dashed horizontal line indicating the Fermi level $E_F$. The band structures and DOS near $E_F$ enable comparison of electronic band gaps, exchange splitting, and spectral weight, providing a systematic assessment of the magnetic and insulating characteristics of these candidate materials.}
\label{4matplot}

\end{figure}

\subsection{First-principles Analysis}
The first-principles calculations were carried out following the spin-polarized DFT workflow illustrated in Fig.~\ref{DFTworkflow}. Starting from the optimized crystal structures, spin-polarized SCF calculations were performed, followed by phonon and electronic-structure analyses.
For both compounds, the phonon dispersions show no imaginary modes across the Brillouin zone (see Fig.~\ref{4matplot}), establishing dynamical stability at $0$ K. According to the workflow shown in Fig.~\ref{DFTworkflow}, the absence of imaginary frequencies confirms dynamical stability after structural relaxation.
The spin-polarized band structures and densities of states display a depletion of spectral weight at the Fermi level (Fig.~\ref{4matplot}), consistent with finite PBE band gaps exceeding 0.025~eV. Pronounced spin asymmetry, reflected in exchange splitting and unequal spin-resolved DOS, indicates a nonzero net magnetic moment. The coexistence of finite band gaps and spin polarization is consistent with exchange-driven insulating states associated with localized transition-metal or rare-earth orbitals. Together with the dynamical stability confirmed through phonon calculations (Fig.~\ref{DFTworkflow}), these first-principles results identify the studied materials as dynamically stable magnetic-insulator candidates.

\section{Discussion}

This work establishes a \emph{constraint-first, property-driven} paradigm for generative materials discovery. Unlike conventional \textit{stability-first} strategies, which generate large candidate sets and subsequently filter for functionality, our approach embeds target properties directly into the generative and selection loop. This enables the search to operate within a highly constrained region of materials space from the outset, improving efficiency in regimes where relevant candidates are intrinsically rare.

Our results demonstrate that this formulation is particularly effective for multi-objective problems defined by competing physical constraints, such as magnetic insulators. Rather than maximizing stability alone, MagMatLLM preferentially generates candidates that simultaneously satisfy stability, magnetism, and insulating behavior, leading to a higher yield of functionally relevant materials per unit computational cost. This shift from post hoc filtering to constraint-guided generation represents a key conceptual advance for materials discovery in data-scarce regimes.

At the same time, several limitations remain. The accuracy of surrogate models ultimately bounds the quality of candidate selection, and improvements in learned interatomic potentials and electronic-structure predictors will directly enhance performance. In addition, while the framework reduces computational cost relative to large-scale screening, the integration of more efficient multi-fidelity evaluation strategies could further improve scalability. Finally, the reliance on existing datasets for initialization introduces potential bias, motivating future work on expanding diversity through more flexible generative priors.

More broadly, the constraint-guided framework introduced here is not limited to magnetic insulators. It provides a general strategy for discovering materials governed by multiple competing criteria, including superconductors, thermoelectrics, and quantum topological phases. By embedding physically interpretable objectives directly into the generative loop, this approach opens a pathway toward controllable, target-aware materials design in regimes where conventional data-driven methods are least effective.

\section*{Conclusion}
This work establishes a \emph{constraint-first} paradigm for generative materials discovery in systems governed by competing physical requirements. By embedding magnetism, insulating behavior, and stability directly into the generative loop, and validating candidates through first-principles calculations, we demonstrate that targeted quantum materials can be identified systematically—even in data-scarce regimes where conventional intuition and high-throughput screening are ineffective. This approach enables direct exploration of materials spaces defined by competing physical constraints, opening immediate pathways to classes that remain largely inaccessible, including magnetic topological insulators, axion insulators, and low-dissipation spin-transport platforms. More broadly, the ability to encode physically interpretable objectives into generative workflows provides a foundation for \emph{target-aware materials design}, where the search process is aligned from the outset with the desired functionality. In this sense, our framework goes beyond accelerating discovery: it introduces a constraint-guided methodology for engineering complex quantum matter, in which generation, evaluation, and physical insight are integrated into a unified, closed-loop design process.

\section*{Acknowledgment}
We acknowledge illuminating discussions with Prashant Padmanabhan.

\section*{Author Contributions}
Q.Z. built the AI model in consultation with T.C. and M.S.H. Q.Z. performed the first-principles calculations in consultation with M.S.H. M.S.H. conceived and supervised the project. All authors contributed to the manuscript preparation and review.

\section*{Funding}
M.S.H. acknowledges support from the Samueli Foundation and The UCLA Council on Research.

\section*{Data Availability}
All data supporting the findings of this study are available within the paper.

\bibliographystyle{naturemag}
\bibliography{sample.bib}

\end{document}